\documentclass[11pt,twoside]{article}
\usepackage{asp2006}
\usepackage{adassconf}

\begin{document}

\paperID{C.10}

\title{Planning and Executing Airborne Astronomy Missions for SOFIA}

\markboth{Gross and Shuping}{Planning and Executing SOFIA Missions}

\author{Michael A. K. Gross and Ralph Y. Shuping}

\affil{Universities Space Research Association,
       NASA/Ames Research Center, MS 211-3,
       Moffett Field, CA 94035 USA}

\contact{Michael A. K. Gross}
\email{mgross@sofia.usra.edu}

\paindex{Gross, M.~A.~K.}
\aindex{Shuping, R.~Y.}


\begin{abstract}
SOFIA is a 2.5 meter airborne infrared telescope, mounted in
a Boeing 747SP aircraft. Due to the large
size of the telescope, only a few degrees of azimuth are available at
the telescope bearing. This means the heading of the aircraft is
fundamentally associated with the telescope's observation targets, and
the ground track necessary to enable a given mission is highly complex
and dependent on the coordinates, duration, and order of observations to
be performed. We have designed and implemented a Flight Management
Infrastructure (FMI) product in order to plan and execute such missions
in the presence of a large number of external constraints (e.g.
restricted airspace, international boundaries, elevation limits of the
telescope, aircraft performance, winds at altitude, and ambient
temperatures).
We present an overview of the FMI, including the process,
constraints and basic algorithms used to plan and execute SOFIA
missions.
\end{abstract}

\section{Introduction}
The Flight Management Infrastructure (FMI) product is intended to
keep the aircraft from interfering with preplanned observations on
the sky.  It predicts the ground tracks necessary to execute its
mission, and corrects the plan for actual conditions while airborne.
To support
this, it contains both a planning component that can run on the ground and in
the air, and an execution component that runs in the air. The planning
component manages a set of ordered observations and optional aircraft
repositioning requests.
The execution component compares the plan to actual
conditions in flight and requests headings (indirectly) from the autopilot.

SOFIA mission planning differs from satellite or ground based
observatory planning in a few key areas.
Most importantly, the observatory position is a function of observation
target history, which prevents observations from being considered as
time-slots alone.  Assignment of flight dates is also nontrivial ---
targets cannot be localized on the
sky or the observatory will always fly in about the same direction
(requiring nearly equivalent dead time to return); this suggests entire
flight series should be considered at once, for greater target variety.

Flight planning and execution differs from conventional as well.
All conventional aircraft fly from point to point along specified
paths on the ground, and ``drift'' the aircraft to compensate for winds.
Typical drift angles (course - heading) exceed 3$^\circ$, and the worst
possible case approaches 30$^\circ$, so SOFIA cannot
necessarily observe in this manner.  Expected winds can be planned for
using a weather forecast.
Correcting course for \emph{unexpected} winds can be accomplished by adjusting
observation durations, by relocating the aircraft between
observations, or by ``observation triage'' as a last resort.  This
requires an astronomically-aware airborne monitoring function to compare
current conditions to plan.
Aircraft capabilities are a strong function of fuel weight, which argues
against simple parametrizing by time, in favor of fuel.

\section{External Constraints}
In addition to the geometrical and practical constraints described
above, the SOFIA flight planning problem also has a number of external
constraints, all of which prevent a truly automated, or even rigorously
sequential, flight planning
process.  For instance, special use airspace (SUA) incursions may require
external approval, and it cannot be known ahead of time whether such
approval will be forthcoming in all cases.
National airspace boundaries require international agreements.
Over-ocean operations are prohibited for safety reasons for the first flights;
for later flights, fairly complex fuel reserve constraints are required.
Gross takeoff weight has a hard limit of 700,000 pounds, which limits
the duration of SOFIA missions.

Some science-driven constraints require interaction with scientists or
detailed knowledge of the observations; especially, trading off water
vapor overburden estimates with altitude and duration, and for trading
off observations against each other.

\section{SOFIA FMI}
Prior to any particular mission, several iterations of flight planning
are performed.  This is expected to include
fully integrated automated flight planning (Frank, Gross, \& K\"urkl\"u
2004), routinely.  Manual choice of observation (including order) will
occur subsequently.  Upon execution, replanning might occur if
conditions are sufficiently different from assumptions.
\begin{figure}[!ht]
\plotone{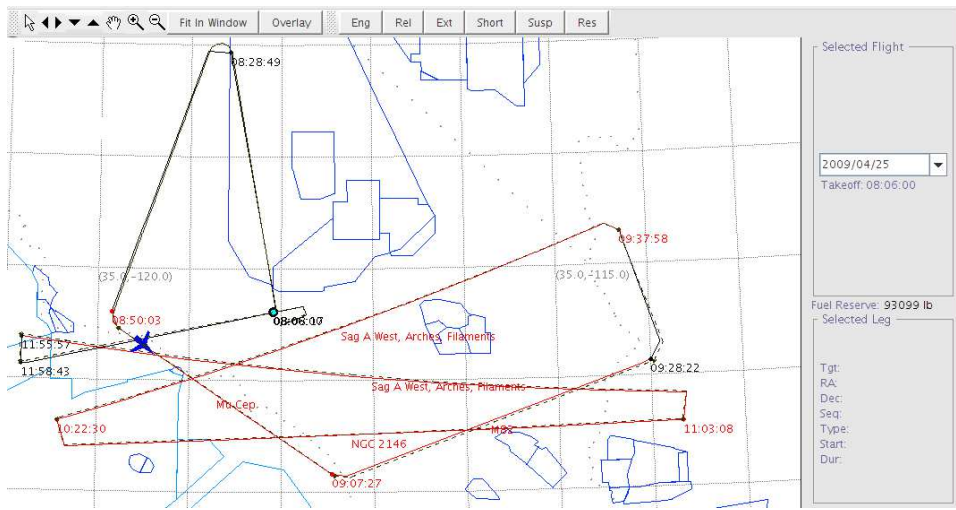}
\caption{A sample flight in southern California airspace, executing in
a simulated environment.  Coastlines and Arizona and Nevada borders are
shown as coarsely dotted lines.
Planned and actual tracks are shown as solid lines.  Since actual
conditions never exactly match plans, the remainder of the flight
is projected from the current actual location, shown as a dashed line.
Offshore polygons are warning zones; others are restricted airspace.
The current position and heading is shown as an aircraft glyph, near
Santa Barbara, CA.}
\label{fig:sampleflight}
\end{figure}
Figure~\ref{fig:sampleflight} shows a color-processed screenshot of a simulated
flight intended for April, 2008, from Palmdale, CA.
Actual conditions
for the simulation shown differ from planned only by small timing errors of
the order of several seconds between segments and at takeoff.

\begin{figure}[!ht]
\plotone{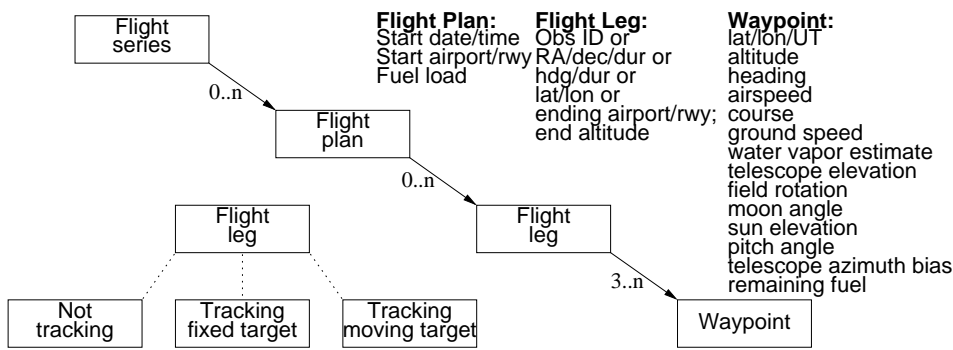}
\caption{Structure of a flight series, corresponding to all
flights between a given instrument's installation and its removal.}
\label{fig:flightseries}
\end{figure}
As mentioned earlier, it is advantageous to consider entire flight
series at once, in order to trade observations between flights.
The data structure
supporting this is shown in Fig.~\ref{fig:flightseries}.

FMI requires substantial input data in order to accurately predict a
flight track and its constraints.  Weather forecast time-series are
taken from the
National Center for Environmental Prediction Global Forecasting System,
quadrilinearly interpolated.  An
alternative is required for dates more than several days in the future,
since accurate forecasts are not available then; we use a set of stacked
monthly means for 1997--2001 (the last years available) from the
European Center for Medium Ranger Weather Forcasting
40 Year Reanalysis (Uppala et. al. 2005) also quadrilinearly interpolated.
Aircraft performance is interpolated from tables generated by Boeing INFLT
runs, for cruising, thrust-limited climbs, and descents.

Planned flight track intersections with special use airspace (SUA) boundaries
(including non-US zones, from
the US National Imagery and Mapping Agency) are evaluated using a quadtree-based search on the 8000+
SUA boundaries for each flight segment.  Observation segments are
treated as initial value problems
in Cartesian coordinates, others may be boundary or initial problems,
as appropriate.  Desired headings
are calculated
during execution from planned (not actual) sky coordinates and actual
position; direct steering by the telescope cannot be allowed for safety
reasons.

In order to test FMI components and integrate other systems, as well as
for training purposes, we use a simulation environment. This includes
a medium-fidelity aircraft simulator, an automated pilot simulator,
a method to set the time arbitrarily, and a telescope simulator
(Br\"uggenwirth, Gross, Nelbach, \& Shuping 2008).

\begin{figure}[!ht]
\plotone{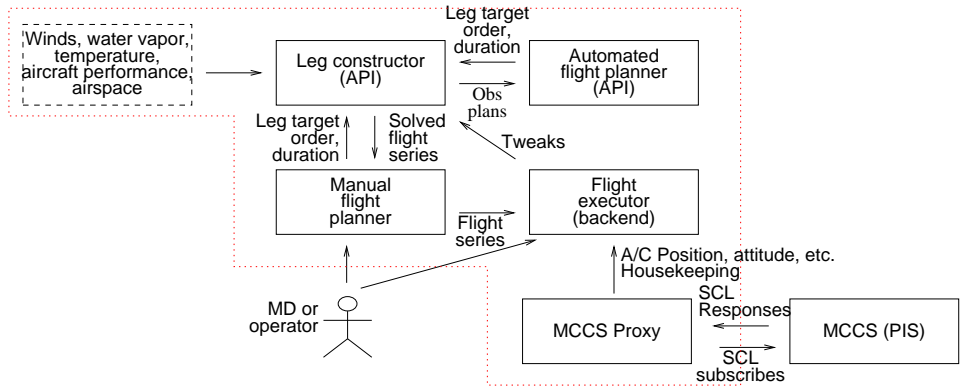}
\caption{Dataflow within the airborne configuration for early flights.
Later flights will have direct connections to the data acquisition
system (MCCS), rather than through a proxy.}
\label{fig:dataflow}
\end{figure}
While airborne, the FMI software components interact with the
airborne data acquisition software to acquire the aircraft's position
and attitude.  While on the ground,
the planner portion of FMI interacts with observers' planning software
and must support multiple simultaneous planning sessions.  The airborne
configuration is shown in Fig.~\ref{fig:dataflow}; the ground
configuration is similar, except there is no flight executor nor MCCS
data (except in testing configurations), and there is a connection to the
Observation Planning Database.

\section{Conclusion}
SOFIA presents a unique flight planning problem due to the nature of
astronomical observation.
FMI provides a connection between scientific needs of an observatory
with the practical constraints of operating an aircraft,
without introducing excessive safety considerations or pilot workload,
or planner effort.

\acknowledgments The authors wish to thank the European Center for
Medium Range Weather Forecasting for access to their reanalysis data.


\begin{references}
\reference Br\"uggenwirth, S., Gross, M.~A.~K., Nelbach, F.~J., \&
  Shuping, R.~Y.\ 2009, \adassxvii, 485
\reference Frank, J., Gross, M.~A.~K., \& K\"urkl\"u, E.\ 2004, in
  Proc. 16th Conf. on Innovative Applications of Artificial Intelligence,
  ed.\ D.~L.\ McGuinness \& G.\ Ferguson (Boston: MIT Press), 828
\reference Uppala, S.~M.\ et al.\ 2005, Quart. J. R. Meterol. Soc.,131, 2961
\end{references}
\end{document}